\documentclass[10pt,a4paper,dvips]{article}
\usepackage[latin1]{inputenc}
\usepackage{graphicx}
\newcommand{\qed}{\hspace*{\fill}$\Diamond$}

\newtheorem{theorem}{Theorem}

\newtheorem{remark}[theorem]{Remark}

\newtheorem{prop}[theorem]{Proposition}
\usepackage{amsmath,amsfonts,amscd,amssymb}
\begin{document}

\title{%%%{\small ARTICLES/antoine/article/2016/ESSAIuniqueVCetDOM.}\\
  On the Complexity of Determining Whether\\
  there is a Unique Hamiltonian Cycle or Path\\
$\:$\\}
\author{
  {\bf Olivier Hudry}\\
  Institut polytechnique de Paris, T\'el\'ecom Paris, 
91123 Palaiseau\\
  %Institut Mines-T\'el\'ecom -- T\'el\'ecom ParisTech\\
  %LTCI, T\'el\'ecom Paris, Universit\'e Paris-Saclay\\
           %19, place Marguerite Perey, 91123 Palaiseau - France\\
%\and
           {\bf \& Antoine Lobstein}\\
           Laboratoire Interdisciplinaire des Sciences du Num\'erique\\
(UMR 9015), CNRS, Universit\'e Paris-Saclay, 91400 Orsay
           %Centre National de la Recherche Scientifique\\
           %Laboratoire de Recherche en Informatique, UMR 8623,\\
           %Universit\'e Paris-Sud, Universit\'e Paris-Saclay\\
           %B\^atiment 650 Ada Lovelace, 91405 Orsay Cedex - France\\
           $\:$\\
           olivier.hudry@telecom-paris.fr, antoine.lobstein@lri.fr   
           %{\bf Olivier Hudry}\\
           % Institut T\'el\'ecom  - T\'el\'ecom ParisTech \& CNRS - LTCI UMR 5141\\
           %46, rue Barrault, 75634 Paris Cedex 13 - France\\
           %hudry@telecom-paristech.fr
%\and
           %{\bf Antoine Lobstein}\\
           %CNRS - LTCI UMR 5141 \&  Institut T\'el\'ecom  - T\'el\'ecom ParisTech\\
           %46, rue Barrault, 75634 Paris Cedex 13 - France\\
           %lobstein@telecom-paristech.fr\\   
$\:$\\}
%\date{}
\maketitle
\begin{abstract} \noindent The decision problems of the existence of a Hamiltonian cycle or of a Hamiltonian path in a given graph, and of the existence of a truth assignment satisfying a given Boolean formula~${\cal C}$, are well-known {\it NP}-complete problems. Here we study the problems of the {\it uniqueness} of a Hamiltonian cycle or path in an undirected, directed or oriented graph, and show that they have the same complexity, up to polynomials, as the problem U-SAT of the uniqueness of an assignment satisfying~${\cal C}$. As a consequence, these Hamiltonian problems are {\it NP}-hard and belong to the class~{\it DP}, like U-SAT. %%%{\bf TS??} We also investigate the closely related problem of the Travelling Salesman {\bf (TS) ..... / GareyJohnson p116 TSE / strongly NP-c pbs}
\end{abstract}

\bigskip

%\pagebreak
%
%$\mbox{ }$
%
%\medskip
%
%\noindent {\bf Running Head:} 
%
%\medskip
%
\pagebreak

\section{Introduction} \label{s1}
\subsection{The Hamiltonian Cycle and Path Problems} \label{ss1}
We shall denote by $G=(V,E)$ a finite, simple, undirected graph with vertex set~$V$ and edge set~$E$, where an {\it edge} between $x\in V$ and $y\in V$ is indifferently denoted by $xy$ or~$yx$. The {\it order} of the graph is its number of vertices,~$|V|$.

If $V=\{v_1, v_2, \ldots, v_n\}$, a {\it Hamiltonian path} ${\cal HP}=<v_{i_1}v_{i_2}\ldots v_{i_n}>$ is an ordering of all the vertices in~$V$, such that $v_{i_j}v_{i_{j+1}}\in E$ for all~$j$, $1\leq j \leq n-1$. The vertices $v_{i_1}$ and $v_{i_n}$ are called the {\it ends} of~${\cal HP}$. A {\it Hamiltonian cycle} is an ordering ${\cal HC}=<v_{i_1}v_{i_2}\ldots v_{i_n}(v_{i_1})>$ of all the vertices in~$V$, such that $v_{i_n}v_{i_1}\in E$ and $v_{i_j}v_{i_{j+1}}\in E$ for all~$j$, $1\leq j \leq n-1$. Note that the same Hamiltonian cycle admits $2n$ representations, e.g., $<v_{i_2}v_{i_3}\ldots v_{i_n}v_{i_1}(v_{i_2})>$ or $<v_{i_n}v_{i_{n-1}}\ldots v_{i_2}v_{i_1}(v_{i_n})>$.

A {\it directed graph} $H=(X,A)$ is defined by its set $X$ of vertices and its set $A$ of directed edges, also called {\it arcs}, an arc being an ordered pair $(x,y)$ of vertices; with this respect, $(x,y)$ and $(y,x)$ are two different arcs and may coexist. A directed graph is said to be {\it oriented} if it is antisymmetric, i.e., if we have, for any pair $\{x,y\}$ of vertices, at most one of the two arcs $(x,y)$ or~$(y,x)$; if $(x,y)\in A$, we say that $y$ is the {\it out-neighbour} of~$x$, and $x$ is the {\it in-neighbour} of~$y$, and we define the {\it in-degree} and {\it out-degree} of a vertex accordingly. The notions of {\it directed Hamiltonian cycle} and of {\it directed Hamiltonian path} are extended to a directed graph by considering the arcs $(v_{i_n},v_{i_1})\in A$ and $(v_{i_j},v_{i_{j+1}})\in A$ in the above definitions. When there is no ambiguity, we shall often drop the words ``directed'' and ``Hamiltonian''.

%%A (directed or undirected) structure is said to be {\it Hamiltonian} if it goes through every vertex of the (directed or undirected) graph exactly once. 
The following six problems (stated as one) are well known, in graph theory as well as in complexity theory:

\medskip

\noindent {\bf Problem} HAMC / HAMP (Hamiltonian Cycle / Hamiltonian Path):\\
{\bf Instance:} An undirected, directed or oriented graph.\\
{\bf Question}: Does the graph admit a Hamiltonian cycle / Hamiltonian path?

\medskip

\noindent As we shall see (Proposition~\ref{gahuHAMC}), they have been known to be {\it NP}-complete for a long time. In this paper, we shall be interested in the following problems, and shall locate them in the complexity classes:

\medskip

\noindent {\bf Problem} U-HAMC[U] (Unique Hamiltonian Cycle in an Undirected graph):\\
{\bf Instance:} An undirected graph $G=(V,E)$.\\
{\bf Question}: Does $G$ admit a {\it unique} Hamiltonian cycle?

\medskip

\noindent {\bf Problem} U-HAMP[U] (Unique Hamiltonian Path in an Undirected graph):\\
{\bf Instance:} An undirected graph $G = (V,E)$.\\
{\bf Question:} Does $G$ admit a {\it unique} Hamiltonian path?

\medskip

\noindent {\bf Problem} U-HAMC[D] (Unique directed Hamiltonian Cycle in a Directed graph):\\
{\bf Instance:} A directed graph $H=(X,A)$.\\
{\bf Question:} Does $H$ admit a {\it unique} directed Hamiltonian cycle?

\medskip

\noindent {\bf Problem} U-HAMP[D] (Unique directed Hamiltonian Path in a Directed graph):\\
{\bf Instance:} A directed graph $H=(X,A)$.\\
{\bf Question:} Does $H$ admit a {\it unique} directed Hamiltonian path?

\medskip

\noindent {\bf Problem} U-HAMC[O] (Unique directed Hamiltonian Cycle in an Oriented graph):\\
{\bf Instance:} An oriented graph $H=(X,A)$.\\
{\bf Question:} Does $H$ admit a {\it unique} directed Hamiltonian cycle?

\medskip

\noindent {\bf Problem} U-HAMP[O] (Unique directed Hamiltonian Path in an Oriented graph):\\
{\bf Instance:} An oriented graph $H=(X,A)$.\\
{\bf Question:} Does $H$ admit a {\it unique} directed Hamiltonian path?

\medskip
%
%\noindent %{\bf annoncer les resultats ici / Forthcoming papers}
%
%
\noindent We shall prove in Section~\ref{sec2} that these problems have the same complexity, up to polynomials, as the problem of the uniqueness of a truth assignment satisfying a Boolean formula (U-SAT). As a consequence, all are {\it NP}-hard and belong to the class~{\it DP}. %%%{\bf TS?? We also investigate the closely related problem of the Travelling Salesman (TS) .....}
The closely related problem Unique Optimal Travelling Salesman has been investigated in~\cite{papa84a}, see Remark~\ref{pap}.

In similar works, we reexamine some famous problems, from the viewpoint of uniqueness of solution: Vertex Cover and Dominating Set (as well as its generalization to domination within distance~$r$)~\cite{hudr16b}, $r$-Identifying Code together with $r$-Locating-Dominating Code~\cite{hudr16d}, and Graph Colouring and Boolean Satisfiability~\cite{hudr16a}. We shall re-use here results from~\cite{hudr16a}, and modify a construction from~\cite{hudr16b}.

\medskip

\noindent In the sequel, we shall need the following tools, which constitute classical definitions related to graph theory or to Boolean satisfiability. A {\it vertex cover} in an undirected graph~$G$ is a subset of vertices $V^*\subseteq V$ such that for every edge $e=uv\in E$, $V^*\cap \{u,v\} \neq \emptyset$. We denote by $\phi(G)$ the smallest cardinality of a vertex cover of~$G$; any vertex cover~$V^*$ with $|V^*|=\phi(G)$ is said to be {\it optimal}.

Next we consider a set ${\cal X}$ of $n$ {\it Boolean variables} $x_i$ and a set ${\cal C}$ of $m$ {\it clauses} (${\cal C}$~is also called a {\it Boolean formula}); each clause $c_j$ contains $\kappa_j$ {\it literals}, a literal being a variable~$x_i$ or its complement~$\overline{x}_i$. A {\it truth assignment} for~${\cal X}$ sets the variable $x_i$ to TRUE, also denoted by~T, and its complement to FALSE (or~F), or {\it vice-versa.} A truth assignment is said to {\it satisfy} the clause~$c_j$ if $c_j$ contains at least one true literal, and to satisfy the set of clauses~${\cal C}$ if every clause contains at least one true literal. The following decision problems %, for which the size of the instance is polynomially linked to $n+m$,
are classical problems in complexity.

\medskip

\noindent {\bf Problem} VC (Vertex Cover with bounded size):\\
{\bf Instance:} An undirected graph $G$ and an integer $k$.\\
{\bf Question}: Does $G$ admit a vertex cover of size at most~$k$?

\medskip

\noindent {\bf Problem} SAT (Satisfiability):\\
{\bf Instance:} A set ${\cal X}$ of variables, a collection ${\cal C}$ of clauses over~${\cal X}$, each clause containing at least two different literals.\\
{\bf Question}: Is there a truth assignment for~${\cal X}$ that satisfies~${\cal C}$?

\medskip

\noindent The following problem is stated for any fixed integer $k\geq2$.

\medskip

\noindent {\bf Problem} $k$-SAT ($k$-Satisfiability):\\
{\bf Instance:} A set ${\cal X}$ of variables, a collection ${\cal C}$ of clauses over~${\cal X}$, each clause containing exactly $k$ different literals.\\
{\bf Question}: Is there a truth assignment for~${\cal X}$ that satisfies~${\cal C}$?

%\medskip
%
%\noindent {\bf Problem} 3-SAT (Three-Satisfiability):\\
%{\bf Instance:} A set ${\cal X}$ of variables, a collection ${\cal C}$ of clauses over~${\cal X}$, each clause containing exactly three different literals.\\
%{\bf Question}: Is there a truth assignment for~${\cal X}$ that satisfies~${\cal C}$?
%
\medskip

\noindent {\bf Problem} 1-3-SAT (One-in-Three Satisfiability):\\
{\bf Instance:} A set ${\cal X}$ of variables, a collection ${\cal C}$ of clauses over~${\cal X}$, each clause containing exactly three different literals.\\
{\bf Question}: Is there a truth assignment for~${\cal X}$ such that each clause of~${\cal C}$ contains {\it exactly one} true literal?

\medskip

\noindent We shall say that a clause (respectively, a set of clauses) is 1-3-{\it satisfied} by an assignment if this clause (respectively, every clause in the set) contains exactly one true literal. 
%
%\medskip
%
%\noindent {\bf Problem} co-SAT (co-Satisfiability):\\
%{\bf Instance:} A set ${\cal X}$ of variables, a collection ${\cal C}$ of clauses over~${\cal X}$, each clause containing at least two different literals.\\
%{\bf Question}: Is it true that no truth assignment for~${\cal X}$ satisfies~${\cal C}$?
%
%\medskip
%
%\noindent
We shall also consider the following variants of the above problems: 

U-VC (Unique Vertex Cover with bounded size),

U-SAT (Unique Satisfiability), 

U-$k$-SAT (Unique $k$-Satisfiability),

U-1-3-SAT (Unique One-in-Three Satisfiability).

\noindent They have the same instances as VC, SAT, $k$-SAT and 1-3-SAT respectively, but now the question is ``Is there a {\it unique} vertex cover / truth assignment$\ldots$?''. 

We shall give in Propositions~\ref{propALOH1}--\ref{propALOH4} what we need to know about the complexities of these problems. 

\subsection{Some Classes of Complexity} \label{ss2}
We refer the reader to, e.g., \cite{bart96}, \cite{gare79}, \cite{john90} or \cite{papa94} for more on this topic. A {\it decision problem} is of the type ``Given an instance~$I$ and a property~${\cal PR}$ on~$I$, is ${\cal PR}$ true for~$I$?'', and has only two solutions, ``yes'' or ``no''. The class {\it P} will denote the set of problems which can be solved by a {\it polynomial} (time) algorithm, and the class {\it NP} the set of problems which can be solved by a {\it nondeterministic polynomial} algorithm. A {\it polynomial reduction} from a decision problem~$\pi_1$ to a decision problem~$\pi_2$ is a polynomial transformation that maps any instance of~$\pi_1$ into an equivalent instance of~$\pi_2$, that is, an instance of~$\pi_2$ admitting the same answer as the instance of~$\pi_1$; in this case, we shall write $\pi_1 \leqslant_p \pi_2$. Cook~\cite{cook71} proved that there is one problem in {\it NP}, namely SAT, to which every other problem in {\it NP} can be polynomially reduced. Thus, in a sense, SAT is the ``hardest'' problem inside~{\it NP}. Other problems share this property in {\it NP} and are called {\it NP}-{\it complete} problems; their class is denoted by {\it NP}-{\it C}. The way to show that a decision problem $\pi$ is {\it NP}-complete is, once it is proved to be in {\it NP}, to choose some {\it NP}-complete problem $\pi_1$ and to polynomially reduce it to~$\pi$. {F}rom a practical viewpoint, the {\it NP}-completeness of a problem $\pi$ implies that we do not know any polynomial algorithm solving~$\pi$, and that, under the assumption {\it P}$\, \neq \,${\it NP}, which is widely believed to be true, no such algorithm exists: the time required can grow exponentially with the size of the instance (when the instance is a graph, its size is polynomially linked to its order; for a Boolean formula, the size is polynomially linked to, e.g., the number of variables plus the number of clauses).

The {\it complement} of a decision problem, ``Given~$I$ and~${\cal PR}$, is ${\cal PR}$ true for~$I$?'', is ``Given~$I$ and~${\cal PR}$, is ${\cal PR}$ false for~$I$?''. The class {\it co}-{\it NP} (respectively, {\it co}-{\it NP}-{\it C}) is the class of the problems which are the complement of a problem in {\it NP} (respectively, {\it NP}-{\it C}).

For problems which are not necessarily decision problems, a {\it Turing reduction} from a problem $\pi_1$ to a problem $\pi_2$ is an algorithm ${\cal A}$ that solves~$\pi_1$ using a (hypothetical) subprogram~${\cal S}$ solving $\pi_2$ such that, if ${\cal S}$ were a polynomial algorithm for $\pi_2$, then ${\cal A}$ would be a polynomial algorithm for~$\pi_1$. Thus, in this sense, $\pi_2$ is ``at least as hard'' as~$\pi_1$. A problem $\pi$ is {\it NP}-{\it hard} (respectively, {\it co}-{\it NP}-{\it hard}) if there is a Turing reduction from some {\it NP}-complete (respectively, co-{\it NP}-complete) problem to~$\pi$ \cite[p.~113]{gare79}. 
\begin{remark} \label{rgj79}
Note that with these definitions, {\it NP}-hard and co-{\it NP}-hard coincide~\cite[p.~114]{gare79}.
\end{remark}
The notion of completeness %and hardness 
can of course be extended to classes other than {\it NP} or co-{\it NP}. Observe that {\it NP}-hardness is defined differently in~\cite{corm00} and~\cite{hema00}: there, a problem $\pi$ is {\it NP}-{\it hard} if there is a {\it polynomial} reduction from some {\it NP}-complete problem to~$\pi$; this may lead to confusion (see Section~\ref{secConc}).

\begin{figure}
\begin{center}
\includegraphics*[scale=1.2]{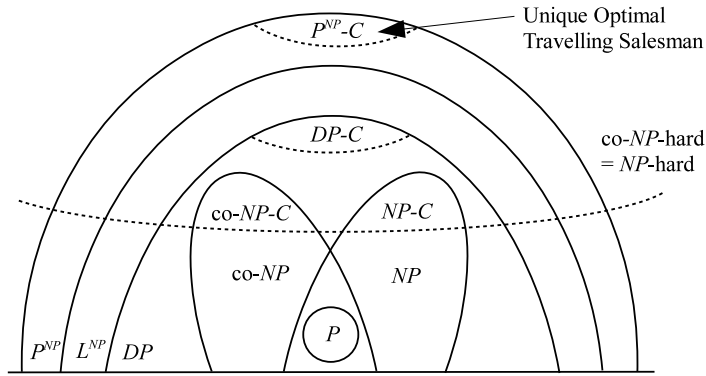}
\end{center}
\caption{Some classes of complexity.}
\label{f1Lh}
\end{figure}

We also introduce the classes {\it P}$^{NP}$ (also known as $\Delta_2$ in the hierarchy of classes) and {\it L}$^{NP}$ (also denoted by {\it P}$^{NP[O(\log n)]}$ or~$\Theta_2$), which contain the decision problems which can be solved by applying, with a number of calls which is polynomial (respectively, logarithmic) with respect to the size of the instance, a subprogram able to solve an appropriate problem in {\it NP} (usually, an {\it NP}-complete problem); and the class {\it DP}~\cite{papa84} (or {\it DIF}$^P$~\cite{blas82} or {\it BH}$_2$~\cite{john90}, \cite{zoo}$,~\ldots$) as the class of languages (or problems)~$L$ such that there are two languages $L_1\in\,${\it NP} and $L_2\in\,$co-{\it NP} satisfying $L=L_1\cap L_2$ (in Figure~\ref{f1Lh}, \textit{DP-C} and $P^{NP}$-\textit{C} denote respectively the class of the \textit{DP}-complete problems and the class of the $P^{NP}$-complete problems). This class is not to be confused with {\it NP}$\,\cap\,$co-{\it NP} (see the warning in, e.g.,~\cite[p.~412]{papa94}); actually, {\it DP} contains {\it NP}$\,\cup\,$co-{\it NP} and is contained in {\it L}$^{NP}$. See Figure~\ref{f1Lh}.

Membership to {\it P}, {\it NP}, co-{\it NP}, {\it DP}, {\it L}$^{NP}$ or {\it P}$^{NP}$ gives an upper bound on the complexity of a problem (this problem is not more difficult than~$\ldots$), whereas a \textit{NP}-hardness result gives a lower bound (this problem is at least as difficult as any problem belonging to \textit{NP} or to co-\textit{NP}). Still, such results are conditional in the sense that we do not know whether or where the classes of complexity collapse.

We now consider some of the problems from Section~\ref{ss1}.          
\begin{prop} \label{gahuHAMC} \cite{karp72}, \cite[pp. 56--60 and pp. 199-200]{gare79} The decision problems HAMC and HAMP, in an undirected, directed or oriented graph, are {\it NP}-complete. \qed
\end{prop}
The problems VC, SAT and 3-SAT are also three of the basic and most well-known {\it NP}-complete problems~\cite{cook71}, \cite[p.~39, p.~46, p.~190 and p.~259]{gare79}. %It follows that the problem co-SAT is co-{\it NP}-complete.
More generally, $k$-SAT is {\it NP}-complete for $k\geq3$ and polynomial for $k=2$. The problem 1-3-SAT, which is obvioulsy in~{\it NP}, is also {\it NP}-complete \cite[Lemma~3.5]{scha78}, \cite[p.~259]{gare79}, \cite[Rem.~3]{hudr16a}.

The following results will be used in the sequel.
\begin{prop} \label{propALOH1}
  \cite{hudr16a} For every integer $k\geq 3$, the decision problems U-SAT, U-$k$-SAT and U-$1$-$3$-SAT have equivalent complexity, up to polynomials. \qed
\end{prop}
Using the previous proposition and results from~\cite{blas82} and \cite[p.~415]{papa94}, it is rather simple to obtain the following two results.

\begin{prop} \label{propALOH2}
For every integer $k\geq 3$, the decision problems U-SAT, U-$k$-SAT and U-$1$-$3$-SAT are co-NP-hard and thus NP-hard by Remark~\ref{rgj79}. \qed

\end{prop}
\begin{prop} \label{propALOH3}
For every integer $k\geq 3$, the decision problems U-SAT, U-$k$-SAT and U-$1$-$3$-SAT belong to the class~DP. \qed

\end{prop}

\begin{remark} \label{rempapa}
It is not known whether these problems are {\it DP}-complete. In \cite[p.~415]{papa94}, it is said that ``U-SAT is not believed to be {\it DP}-complete''.
\end{remark}

\begin{prop} \label{propALOH4} \cite{hudr16b}
The decision problems U-SAT and U-VC have equivalent complexity, up to polynomials. In particular, there exists a polynomial reduction from U-$1$-$3$-SAT to U-VC: U-$1$-$3$-SAT $\leqslant_p$ U-VC. \qed
  \end{prop}
  
After the following remark %s are made
is made, we shall be ready to investigate the problems of uniqueness of Hamiltonian cycle or path.
\begin{remark} \label{pap}
  In \cite{papa84a}, it is shown that the following problem is {\it P}$^{NP}$-complete (or $\Delta_2$-complete).
  
  \medskip

\noindent {\bf Problem} U-OTS (Unique Optimal Travelling Salesman):\\
{\bf Instance:} A set of $n$ vertices, a $n\times n$ symmetric matrix $[c_{ij}]$ of (nonnegative) integers giving the distance between any two vertices $i$ and~$j$.\\
{\bf Question}: Is there a unique optimal tour, that is, a unique way of visiting every vertex exactly once and coming back, with the smallest distance sum?

  \end{remark}
%\medskip

\noindent At best, a polynomial reduction from any instance $G=(V,E)$ of U-HAMC[U] to U-OTS would show that U-HAMC[U] belongs to {\it P}$^{NP}$, but we have a better result in Theorem~\ref{gros}(b), with U-HAMC[U] belonging to~{\it DP}; no useful information for our Hamiltonian problems can be induced from this result on U-OTS.
%Since there is an easy polynomial reduction from any instance $G=(V,E)$ of U-HAMC[U] to U-OTS (simply take $n=|V|$ and $c_{ij}=1$ if $ij\in E$, $c_{ij}=2$ otherwise), this shows that U-HAMC[U] belongs to {\it P}$^{NP}$, but we have a better result in Theorem~\ref{gros}(b) with U-HAMC[U] belonging to~{\it DP}.
%However, no useful information for our problems can be directly induced from this result on U-OTS.
%\begin{remark}
%   In \cite{vali86}, it is mentioned that \cite{vali74} gives a ``reduction from satisfiability to Hamiltonian circuits that preserves the number of solutions''; this would be of course stronger than our Theorem~\ref{13SATHAMC} which gives a reduction from U-1-3-SAT to U-HAMC, and its consequence, Theorem~\ref{gros}(a), but \cite{vali74} is an unpublished manuscript. And we also give a reduction from U-HAMC to U-SAT (Theorem~\ref{UHAMU3SAT}), which shows the equivalence of the two problems.
%  \end{remark}
\section{Locating the Problems of Uniqueness} \label{sec2}
\begin{figure}
\begin{center}
\includegraphics*[scale=0.82]{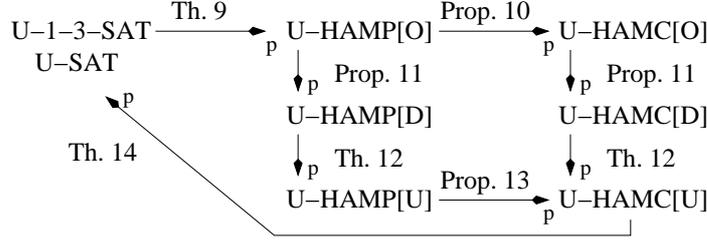}
\end{center}
\caption{The chain of polynomial reductions, where an arrow from $\pi_1$ to $\pi_2$ stands for the relation $\pi_1 \leqslant _p \pi_2$.}
\label{chainPR}
\end{figure}

We prove that our six Hamiltonian problems have the same complexity as any of the three problems U-SAT, U-$k$-SAT ($k\geq 3$) and U-1-3-SAT by proving the chain of polynomial reductions given by Figure~\ref{chainPR}.

%%%U-1-3-SAT $\leqslant_p$ U-HAMC (Theorem~\ref{13SATHAMC}) and U-HAMC $\leqslant_p$ U-SAT (Theorem~\ref{UHAMU3SAT}).
\begin{theorem} \label{13SATHAMC}
There exists a polynomial reduction from U-$1$-$3$-SAT to U-HAMP[O]: U-$1$-$3$-SAT $\leqslant_p$ U-HAMP[O].
%The decision problem U-HAMC is at least as difficult as the problem U-1-3-SAT.
\end{theorem}
\noindent {\bf Proof.}  %%% We describe a polynomial reduction from the problem U-1-3-SAT to U-HAMP[O], via U-VC: to do this, we use a polynomial reduction from U-1-3-SAT to U-VC, which is simpler than the one given in the proof of Theorem~8 in~\cite{hudr16b} and mentioned in Proposition~\ref{propALOH4} of this paper; that proof had the advantage to cope with U-VC and U-OVC (Unique Optimal Vertex Cover) at the same time, but would give here a much more complicated proof for U-HAMP[O].
A polynomial reduction from the problem U-1-3-SAT to U-HAMP[O], via U-VC, can be found in the Appendix; it is an elaborate variation on the polynomial reduction from 3-SAT to VC in~\cite{karp72}, \cite[pp. 54--56]{gare79} and the polynomial reduction from VC to HAMC[U] (see~\cite[pp. 56--60]{gare79}). \qed

\begin{prop} \label{theoPOCO}
  There exists a polynomial reduction from U-HAMP[O] to U-HAMC[O]: U-HAMP[O] $\leqslant_p$ U-HAMC[O].
\end{prop}
\noindent {\bf Proof.} We start from an oriented graph $H=(X,A)$ which is an instance of U-HAMP[O] and build a graph which is an instance of U-HAMC[O] by adding two extra vertices~$y, \,z$, together with the arc $(y,z)$ and all the arcs $(x,y)$ and $(z,x)$, $x\in X$. This transformation is polynomial and clearly preserves the number of solutions, in particular the uniqueness. \qed

\begin{prop} \label{propPDCD}
  There is a polynomial reduction from U-HAMP[O] to U-HAMP[D] and from U-HAMC[O] to U-HAMC[D]:

  U-HAMP[O] $\leqslant _p$ U-HAMP[D] and U-HAMC[O] $\leqslant _p$ U-HAMC[D].
\end{prop}
\noindent {\bf Proof.} It suffices to consider the identity as the polynomial reduction. \qed
\begin{theorem} \label{triplica}
   There is a polynomial reduction from U-HAMP[D] to U-HAMP[U] and from U-HAMC[D] to U-HAMC[U]:

  U-HAMP[D] $\leqslant _p$ U-HAMP[U] and U-HAMC[D] $\leqslant _p$ U-HAMC[U].
\end{theorem}
\noindent {\bf Proof.} The method is borrowed from~\cite{karp72}.

Consider any instance of U-HAMP[D] or U-HAMC[D], i.e., a directed graph $H=(X,A)$ on $n$ vertices. We build the undirected graph $G=(V,E)$, the instance of U-HAMP[U] or U-HAMC[U], as follows: every vertex $x \in X$ is triplicated into three vertices $x^-\in V$ (a {\it minus-type vertex}), $x^*\in V$ (a {\it star-type vertex}) and $x^+\in V$, linked by the edges $x^-x^*\in E$ and $x^*x^+\in E$; for every arc $(x,y)\in A$, we create the edge $x^+y^-$ in~$E$. The graph $G$ thus constructed has order $3n$.

We claim that there is a unique Hamiltonian cycle (respectively, path) in~$G$ if and only if there is a unique directed Hamiltonian cycle (respectively, path) in~$H$.

(1) Assume first that $H$ admits a directed Hamiltonian cycle $<x_1x_2\ldots$ $x_n(x_1)>$. Then 
$$<x_1^-x^*_1x_1^+x_2^-x^*_2x_2^+\ldots x_{n-1}^+x_n^-x^*_nx_n^+(x_1^-)>$$
is a Hamiltonian cycle in~$G$. Moreover, two different directed Hamiltonian cycles in~$H$ provide two different Hamiltonian cycles in~$G$. 

Conversely, assume that $G$ admits a Hamiltonian cycle~${\cal HC}$. This cycle must go through all the star-type vertices $x^*$, so it necessarily goes through all the edges $x^-x^*$ and $x^*x^+$. Without loss of generality, ${\cal HC}$ reads: 
\begin{equation} \label{eqCHCH} 
{\cal HC}=<x_1^-x^*_1x_1^+x_2^-x^*_2x_2^+\ldots x_{n-1}^+x_n^-x^*_nx_n^+(x_1^-)>\,;
\end{equation}
indeed, we may assume that we ``start'' with the edge $x_1^-x_1^*$, then $x_1^*x_1^+$; now, because the edges which have no star-type vertex as one of their extremities are necessarily of the type $x^+y^-$, the other neighbour of $x_1^+$ is a minus-type vertex, say $x_2^-$; step by step, we see that ${\cal HC}$ has necessarily the previous form~(\ref{eqCHCH}). Now we claim that $<x_1x_2\ldots x_{n-1}x_n(x_1)>$ is a directed Hamiltonian cycle in~$H$.

Indeed, for every $i\in \{1, \ldots ,n-1\}$, the edge $x_i^+x_{i+1}^-$ in~$G$ implies the existence of the arc $(x_i,x_{i+1})$ in~$H$; the same is true for the arc $(x_n,x_1)$ in~$H$, thanks to the edge $x_n^+x_1^-$ in~$G$. Furthermore, observe that two different Hamiltonian cycles in~$G$ provide two different directed Hamiltonian cycles in~$H$.

So, $G$ admits a unique Hamiltonian cycle if and only if $H$ admits a unique directed Hamiltonian cycle. 

(2) Exactly the same argument works with paths, apart from the fact that we need not consider the arc $(x_n,x_1)$ in~$H$, nor the edge $x_n^+x_1^-$ in~$G$. \qed
\begin{prop} \label{Uuniv}
    There exists a polynomial reduction from U-HAMP[U] to U-HAMC[U]: U-HAMP[U] $\leqslant_p$ U-HAMC[U].
\end{prop}
\noindent {\bf Proof.} We start from an undirected graph $G=(V,E)$ which is an instance of U-HAMP[U] and build a graph which is an instance of U-HAMC[U] by adding the extra vertex~$y$, together with all the edges $xy$, $x\in V$. This transformation is polynomial and clearly preserves the number of solutions, in particular the uniqueness. \qed
\begin{theorem} \label{UHAMU3SAT}
%The decision problem U-OVC is at least as difficult as the problem U-1-3-SAT.
  There exists a polynomial reduction from U-HAMC[U] to U-SAT: U-HAMC[U] $\leqslant_p$ U-SAT.
\end{theorem}
\noindent {\bf Proof.} We start from an instance of U-HAMC[U], an undirected graph $G=(V,E)$ with $V=\{x^1, \ldots, x^{|V|}\}$; we assume that $|V|\geq 3$. We create the set of variables ${\cal X}=\{x^i_j: 1\leq j \leq |V|, 1\leq i \leq |V|\}$  and the following clauses:

\medskip

(a$_1$) for $1 \leq i \leq |V|$, clauses of size~$|V|$: $\{x^i_1, x^i_2, \ldots, x^i_{|V|}\}$;

(a$_2$) for $1 \leq i \leq |V|$, $1 \leq j<j' \leq |V|$, clauses of size two: $\{{\overline x}^i_j, {\overline x}^i_{j'}\}$;

(b$_1$) for $1 \leq j \leq |V|$, clauses of size~$|V|$: $\{x_j^1, x_j^2, \ldots, x_j^{|V|}\}$;

(b$_2$) for $1 \leq i<i' \leq |V|$, $1 \leq j \leq |V|$, clauses of size two: $\{{\overline x}^i_j, {\overline x}^{i'}_j\}$;

(c) for $1\leq i<i' \leq |V|$ such that $x^ix^{i'} \notin E$, for $1\leq j \leq|V|$, clauses of size two: $\{{\overline x}^i_j, {\overline x}^{i'}_{j+1}\}$ and $\{{\overline x}^i_j, {\overline x}^{i'}_{j-1}\}$, with computations performed modulo~$|V|$;

%+ des clauses pour l'unicite, pour casser les permutations, ou plutot l'orientation  LE MIROIR! ? oui ($x^Â¹$ est le premier mais apres le 2eme peut etre $x^2$ ou $x^{|V|}$
%
(d$_1$) $\{x^1_1\}$;

(d$_2$) for $2\leq j<j' \leq |V|$, clauses of size two: $\{{\overline x}^2_{j'}, {\overline x}^3_j\}$. %{\bf A VOIR}

\medskip

\noindent Assume that we have a unique Hamiltonian cycle in~$G$, ${\cal HC}_1=<x^{p_1}x^{p_2}x^{p_3}$ $\ldots x^{p_{|V|-1}}x^{p_{|V|}}(x^{p_1})>$. Note that for the time being, we could also write ${\cal HC}_1=<x^{p_1}x^{p_{|V|}}x^{p_{|V|-1}}\ldots x^{p_3}x^{p_2}(x^{p_1})>$, or ``start'' on a vertex other than $x^{p_1}$, cf. Introduction. This is why, without loss of generality, we set $p_1=1$, i.e., we ``fix'' the first vertex, and we also choose the ``direction'' of the cycle, by deciding, e.g., that $x^2$ appears ``before'' $x^3$ in the cycle ---cf. (d$_1$)-(d$_2$). Define the assignment ${\cal A}_1$ by ${\cal A}_1(x^{p_q}_q)=\,$T for $1\leq q \leq |V|$, and all the other variables are set FALSE by~${\cal A}_1$. We claim that ${\cal A}_1$ satisfies all the clauses.

(a$_1$) for $1\leq i \leq |V|$, if the vertex $x^i$ has position~$j$ in the cycle, then the variable~$x^i_j$ satisfies the clause; (a$_2$)~if $\{{\overline x}^i_j, {\overline x}^i_{j'}\}$ is not satisfied by~${\cal A}_1$ for some $i,j,j'$, then ${\cal A}_1(x^i_j)= {\cal A}_1(x^i_{j'})=\,$T, which means that the vertex~$x^i$ appears at least twice in the cycle;

(b$_1$) for $1\leq j \leq |V|$, if the position~$j$ is occupied by the vertex~$x^i$, then the variable~$x^i_j$ satisfies the clause; (b$_2$)~if $\{{\overline x}^i_j, {\overline x}^{i'}_j\}$ is not satisfied by~${\cal A}_1$ for some $i,i',j$, then two different vertices are the $j$-th vertex in the cycle.

(c) If one of the two clauses is not satisfied, say the first one, then the positions~$j$ and~$j+1$ in the cycle are occupied by two vertices not linked by any edge in~$G$.

(d$_1$) $\{x^1_1\}$ is satisfied by ${\cal A}_1$ thanks to the assumption on the first vertex of the cycle; (d$_2$)~if for some $j<j'$, the clause $\{{\overline x}^2_{j'}, {\overline x}^3_j\}$ is not satisfied, then the vertex~$x^3$ occupies a position~$j$ smaller than the position~$j'$ of~$x^2$, which contradicts our assumption on~$x^2$ and~$x^3$.

Is ${\cal A}_1$ unique? Assume on the contrary that another assignment,~${\cal A}_2$, also satisfies the constructed instance of U-SAT. Then by~(a$_1$) and~(a$_2$), for every $i\in \{1, \ldots, |V|\}$, there is at least, then at most, one $j=j(i)$ such that ${\cal A}_2(x^i_j)=\,$T; by~(b$_1$) and~(b$_2$), for every $j\in \{1, \ldots, |V|\}$, there is at least, then at most, one $i=i(j)$ such that ${\cal A}_2(x^i_j)=\,$T; so we have ``a place for everything and everything in its place'', with exactly $|V|$ variables which are TRUE by~${\cal A}_2$ and an ordering of the vertices according to the one-to-one correspondence given by~${\cal A}_2$: the vertex~$x^i$ is in position~$j$ if and only if ${\cal A}_2(x^i_j)=\,$T. Next, thanks to the clauses~(c), two vertices following each other in this ordering, including the last and first ones, are necessarily neighbours, so that this ordering is a Hamiltonian cycle,~${\cal HC}_2$. Since we have assumed the uniqueness of the Hamiltonain cycle~${\cal HC}_1$ in~$G$, the two cycles can differ only by their starting points or their ``directions''. However these differences are ruled out by the clauses~(d$_1$) and~(d$_2$), so that the two cycles coincide vertex to vertex, and ${\cal A}_1={\cal A}_2$. So a YES answer for U-HAMC[U] leads to a YES answer for U-SAT.

Assume now that the answer to U-HAMC[U] is negative. If it is negative because there are at least two Hamiltonian cycles, then we have at least two assignments satisfying the instance of U-SAT: we have seen above how to construct a suitable assignment from a cycle, and different cycles obviously lead to different assignments. If there is no Hamiltonian cycle, then there is no assignment satisfying U-SAT, because such an assignment would give a cycle, as we have seen above with~${\cal A}_2$. So in both cases, a NO answer to U-HAMC[U] implies a NO answer to U-SAT. \qed

\medskip

\noindent Gathering all our previous results, we obtain the following theorem.
\begin{theorem} \label{gros}
  For every integer $k\geq 3$, the decision problems U-SAT, U-$k$-SAT and U-$1$-$3$-SAT have the same complexity as U-HAMP[U], U-HAMC[U], U-HAMP[O], U-HAMC[O], U-HAMP[D], and U-HAMC[D], up to polynomials. Therefore,

  (a) the decision problems U-HAMP[U], U-HAMC[U], U-HAMP[O], U-HAMC[O], U-HAMP[D], and U-HAMC[D] are co-NP-hard and thus NP-hard by Remark~\ref{rgj79};

  (b) the decision problems U-HAMP[U], U-HAMC[U], U-HAMP[O], U-HAMC[O], U-HAMP[D], and U-HAMC[D] belong to the class~DP. \qed
\end{theorem}
Note that the membership to~{\it DP} could have been proved directly.
%
%\bigskip
%
%\noindent UPPER BOUND pour le pb: il est dans DP.
% \section{The Travelling Salesman}
% {\bf ?????}
\section{Conclusion} \label{secConc}
\begin{figure}
\begin{center}
\includegraphics*[scale=1.2]{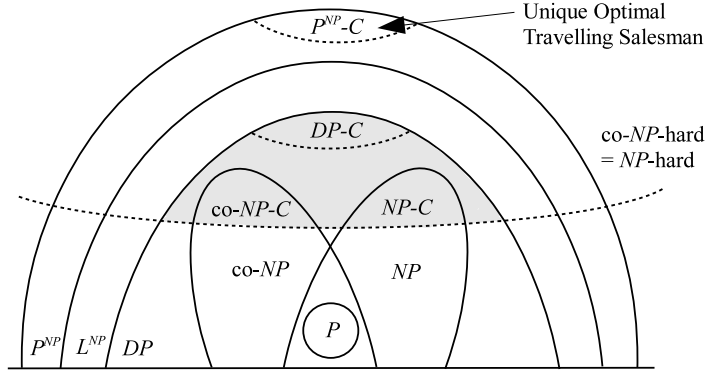}
\end{center}
\caption{Some classes of complexity: Figure~\ref{f1Lh} re-visited.}
\label{dplocrayure}
\end{figure}

By Theorem~\ref{gros}, for every integer $k\geq 3$, the three decision problems U-SAT, U-$k$-SAT, U-$1$-$3$-SAT have the same complexity, up to polynomials, as the problem of the uniqueness of a path or of a cycle in a graph, undirected, directed, or oriented; all are co-{\it NP}-hard (and {\it NP}-hard by Remark~\ref{rgj79}) and belong to the class~{\it DP}, and it is thought that they are not {\it DP}-complete. Anyway, they can be found somewhere in the shaded area of Figure~\ref{dplocrayure}.

\medskip

\noindent {\bf Open problem.} Find a better location for any of these problems inside the hierarchy of complexity classes.

\medskip

\noindent In~\cite{blas82}, the authors wonder whether 

(A) U-SAT is {\it NP}-hard, but here we believe that what they mean is: does there exist a {\it polynomial} reduction from an {\it NP}-complete problem to U-SAT$\,$? i.e., they use the {\it second} definition of {\it NP}-hardness;

finally, they show that (A) is true if and only if

(B) U-SAT is {\it DP}-complete.

\medskip

\noindent So, if one is careless and considers that U-SAT is {\it NP}-hard without checking according to which definition, one might easily jump too hastily to the conclusion that U-SAT is {\it DP}-complete, which, to our knowledge, is not known to be true or not. As for U-3-SAT, we do not know where to locate it more precisely either; in~\cite{cala08} the problems U-$k$-SAT and more particularly U-3-SAT are studied, but it appears that they are versions where the given set of clauses has zero or one solution, which makes quite a difference with our problem.

\bigskip

\noindent {\bf Appendix: the Proof of Theorem~\ref{13SATHAMC}}

\medskip

{\small
\noindent {\bf A) {F}rom U-1-3-SAT to U-VC}\\
{F}rom an arbitrary instance of U-1-3-SAT with $m$ clauses and $n$ variables, we mimick the reduction from 3-SAT to VC in~\cite{karp72}, \cite[pp. 54--56]{gare79}, and we construct the instance $G_{VC}=(V_{VC},E_{VC})$ of U-VC as follows (see Figure~\ref{figWC} for an example): we construct for each clause~$c_j$ a triangle~$T_j=\{a_j,b_j,d_j\}$, and for each variable~$x_i$ a component $G_i=(V_i=\{x_i,\overline{x}_i\},E_i=\{x_i\overline{x}_i\})$. Then we link the components~$G_i$ on the one hand, and the triangles~$T_j$ on the other hand, according to which literals appear in which clauses (``membership edges''). For each clause $c_j=\{\ell_1,\ell_2,\ell_3\}$, we also add the triangular set of edges $E'_j=\{\overline{\ell}_1\overline{\ell}_2, \overline{\ell}_1\overline{\ell}_3, \overline{\ell}_2\overline{\ell}_3\}$. Finally, we set $k=n+2m$.
\begin{figure}
\begin{center}
\includegraphics*[scale=0.85]{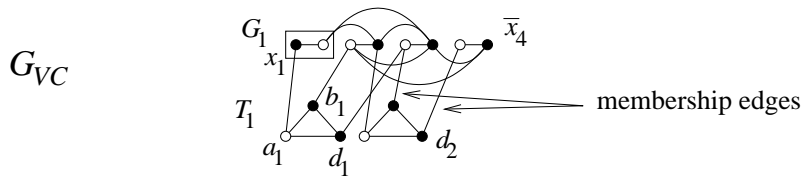}
\end{center}
\caption{Illustration of the undirected graph constructed for the reduction from U-1-3-SAT to U-VC, with four variables and two clauses, $c_1=\{x_1,x_2,x_3\}$, $c_2=\{\overline{x}_2,x_3,x_4\}$. Here, $k=8$, and the black vertices form the (not unique) vertex cover $V^*$ of size eight corresponding to the (not unique) truth assignment $x_1=\,$T, $x_2=\,$F, $x_3=\,$F, $x_4=\,$F 1-3-satisfying the clauses. As soon as we set $V^*\cap (V_1\cup V_2\cup V_3\cup V_4)=\{x_1,\overline{x}_2,\overline{x}_3,\overline{x}_4\}$, the other vertices in $V^*$ are forced.}
\label{figWC}
\end{figure}

The order of $G_{VC}$ is $3m+2n$ and its number of edges is at most $n+9m$ (the edge sets $E'_j$ are not necessarily disjoint).

Note already that if $V^*$ is a vertex cover, then each triangle~$T_j$ contains at least two vertices, each component $G_i$ at least one vertex, and $|V^*|\geq 2m+n=k$; if $|V^*|=2m+n$, then each triangle contains exactly two vertices, and each component $G_i$ exactly one vertex. We can also observe that, because of the edge sets $E'_j$, at least two vertices among $\overline{\ell}_1, \overline{\ell}_2,\overline{\ell}_3$ belong to any vertex cover.

(a) Let us first assume that the answer to U-1-3-SAT is YES: there is a unique truth assignment 1-3-satisfying the clauses of~${\cal C}$. Then, by taking, in each~$G_i$, the vertex corresponding to the literal which is TRUE, and in every triangle~$T_j$, the two vertices which are linked to the two false literals of~$c_j$, we obtain a vertex cover $V^*$ whose size is equal to~$k$. Moreover, once we have put the $n$ vertices corresponding to the true literals in the vertex cover $V^*$ in construction, we have {\it no choice} for the completion of~$V^*$ with $k-n=2m$ vertices: when we take two vertices in~$T_j$, we {\it must} take the two vertices which cover the membership edges linked to the two false literals (in the example of Figure~\ref{figWC}, the vertices $b_1,d_1$ and $b_2,d_2$). So, if another vertex cover $V^+$ of size~$k$ exists, it must have a different distribution of its vertices over the components~$G_i$, still with exactly one vertex in each~$G_i$; this in turn defines a valid truth assignment, by setting $x_i=\,$T if $x_i\in V^+$, $x_i=\,$F if $\overline{x}_i\in V^+$. Now this assignment 1-3-satisfies~${\cal C}$, thanks in particular to our observation on the covering of the edges in $E'_j$. So we have two truth assignments 1-3-satisfying~${\cal C}$, contradicting the YES answer to U-1-3-SAT; therefore, $V^*$ is the only vertex cover of size~$k$.

(b) Assume next that the answer to U-1-3-SAT is NO: this may be either because no truth assignment 1-3-satisfies the instance, or because at least two assignments do; in the latter case, this would lead, using the same argument as in the previous paragraph, to at least two vertex covers of size~$k$, and a NO answer to U-VC. So we are left with the case when the set of clauses~${\cal C}$ cannot be 1-3-satisfied. But again, we have already seen that this would imply that no vertex cover of size (at most)~$k$ exists, since such a hypothetical vertex cover~$V^+$ would imply the existence of a suitable assignment.

We are now ready to construct an instance of U-HAMP[O]. In the sequel, we shall say ``path'' for ``directed Hamiltonian path''.

\medskip

\noindent {\bf B) Construction of the Instance of U-HAMP[O]}\\
We look deeper into the proof of the {\it NP}-completeness of the problem Hamiltonian Cycle (see~\cite[pp. 56--60]{gare79}), which uses a polynomial reduction from VC to HAMC[U] that, due to the so-called ``selector vertices'', cannot cope with the problem of uniqueness; step by step, we construct an oriented graph $H=(X,A)$ for which we will prove that:

(i)~if there is a YES answer for the instance of U-1-3-SAT (which implies that there is a unique vertex cover $V^*$ in $G_{VC}$, with cardinality at most~$k$), then there is a unique path in~$H$;

(ii)~if there are at least two assignments 1-3-satisfying all the clauses (i.e., there are at least two vertex covers in $G_{VC}$, with cardinality at most~$k$), then there are at least two paths in~$H$;

(iii)~if there is no assignment 1-3-satisfying the clauses (and no vertex cover in $G_{VC}$ with cardinality at most~$k$), then there is no path in~$H$. 

\medskip

\noindent {\bf Step 1.} For each edge $e=uv\in E_{VC}$, we build one component $H_e=(X_e,A_e)$ with 12 vertices and 14 arcs: 
 $X_e=\{(u,e,i),(v,e,i): 1\leq i\leq 6\}$, $A_e=\{((u,e,i),(u,e,i+1)), ((v,e,i),(v,e,i+1)): 1\leq i\leq 5\}\cup \{((v,e,3),(u,e,1)),$ $((u,e,3),(v,e,1))\}\cup \{((v,e,6),(u,e,4)),((u,e,6),(v,e,4))\}$; see Figure~\ref{figCyHam1}, which is the oriented copy of Figure~3.4 in~\cite[p.~57]{gare79}.

\begin{figure}
\begin{center}
\includegraphics*[scale=0.9]{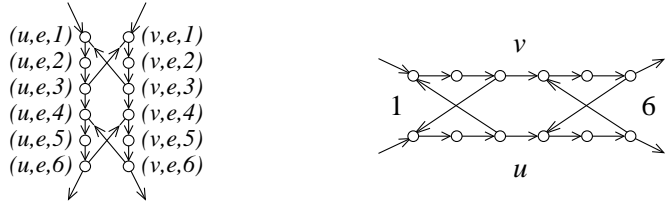}
\end{center}
\caption{Two possible representations of the same component $H_e$ for the edge $e=uv\in E_{VC}$ (Step~1).}
\label{figCyHam1}
\end{figure}
In the completed construction, the only vertices from this component that will be involved in any additional arcs are the vertices $(u,e,1)$, $(u,e,6)$, $(v,e,1)$, and $(v,e,6)$. This, together with the fact that there will be two particular vertices, $\alpha_1$ and~$\delta$, which will necessarily be the ends of any path, will imply that any path in the final graph~$H$ will have to meet the vertices in $X_e$ in exactly one of the three configurations shown in Figure~\ref{figCyHam2}, which is the oriented copy of Figure~3.5, in~\cite[p.~58]{gare79}. Thus, when the path meets the component $H_e$ at $(u,e,1)$, it will have to leave at $(u,e,6)$ and go through either (a)~all 12 vertices in the component, in which case we shall say that the component is {\it completely visited from the} $u$-{\it side}, or (b)~only the 6 vertices $(u,e,i)$, $1 \leq i \leq 6$, in which case we shall say that the component is visited {\it in parallel} and needs two visits, i.e., another section of the path will re-visit the component, meeting the 6 vertices $(v,e,i)$, $1 \leq i \leq 6$. 
\begin{figure}
\begin{center}
\includegraphics*[scale=0.9]{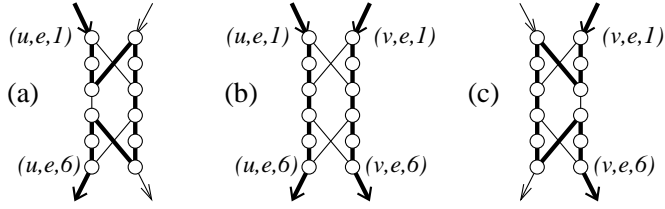}
\end{center}
\caption{The three ways of going through the component $H_e$ (Step~1). The arrows inside $H_e$ are not represented.}
\label{figCyHam2}
\end{figure}

\medskip

\noindent {\bf Step 2.} We create $n$ vertices $\alpha_i$, $1\leq i \leq n$, and $2n$ arcs $(\alpha_i,(x_i,x_i\overline{x}_i,1))$, $(\alpha_i,(\overline{x}_i,x_i\overline{x}_i,1))$, that is, we link $\alpha_i$ to the ``first'' vertices of the component $H_e$ whenever $e=x_i\overline{x}_i$. The vertices $\alpha_i$ can be seen as literal selectors that will choose between $x_i$ and~$\overline{x}_i.$ %We also create the vertex~$\gamma$ and the arc $(\gamma,\alpha_1)$; $\gamma$~will have no other neighbour, so it will necessarily be the starting vertex of any directed Hamiltonian path, if such a path exists.
 The vertex $\alpha_1$ will have no other neighbours; this means in particular that it will have no in-neighbours, thus it will necessarily be the starting vertex of any Hamiltonian path, if such a path exists. 

We choose an arbitrary order on the $3m$ vertices of the triangles $T_j$ in the graph $G_{VC}$, say ${\cal O}_T=<a_1, b_1, d_1, a_2, \ldots, d_m>$ and an arbitrary order on the literals $x_i, \overline{x}_i$, say ${\cal O}_{\ell}=<x_1, x_2, \ldots, x_n, \overline{x}_1, \ldots, \overline{x}_n>$. For each literal $\ell_i$ equal to~$x_i$ or~$\overline{x}_i$, we do the following (see Figure~\ref{figCyHam3} for an example):
\begin{figure}
\begin{center}
\includegraphics*[scale=0.85]{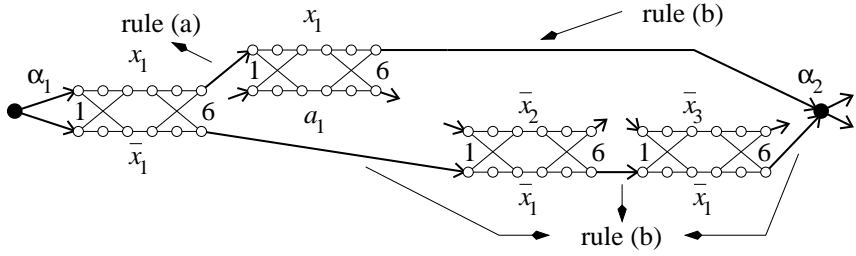}
\end{center}
\caption{The example of the literals $x_1$ and $\overline{x}_1$ from Figure~\ref{figWC} (Step~2); here, Rule~(a) applies with $q(x_1)=1$, $q(\overline{x}_1)=0$, Rule~(b) with $s(\overline{x}_1)=2$. The arrows inside $H_e$ are not represented.}
\label{figCyHam3}
\end{figure}

Rule (a): If $\ell_i$ appears $q=q(\ell_i)\geq 0$ times in the clauses and is linked in~$G_{VC}$ to $t_1, \ldots , t_q$ where the $t$'s belong to the triangles~$T_j$ and follow the order~${\cal O}_T$, then we create the arcs $((\ell_i,\ell_i\overline{\ell}_i,6),(\ell_i,\ell_it_1,1))$, $((\ell_i,\ell_it_1,6),(\ell_i,\ell_it_2,1))$, $\ldots$, $((\ell_i,\ell_it_{q-1},6),(\ell_i,\ell_it_q,1))$.

Rule (b): We consider the triangular sets of edges $E'_j$ described in the construction of~$G_{VC}$.

$\bullet$ If $\ell_i$ does not belong to any such edge, we create the arc $((\ell_i,\ell_it_q,6),\alpha_{i+1})$ ---or $((\ell_i,\ell_i\overline{\ell}_i,6),\alpha_{i+1})$ if $\ell_i$ does not apppear in any clause--- unless $i=n$, in which case we create $((\ell_i,\ell_it_q,6),\beta_1)$ or $((\ell_i,\ell_i\overline{\ell}_i,6),\beta_1)$, where $\beta_1$ is a new vertex that will be spoken of at the beginning of Step~3.

$\bullet$ If $\ell_i$ belongs to $s=s(\ell_i)>0$ edges from $E'_j$, which link $\ell_i$ to $s$ literals $\ell_{i_1}, \ldots ,\ell_{i_s}$ that follow the order ${\cal O}_{\ell}$, then we build the arc $((\ell_i,\ell_it_q,6),(\ell_i,\ell_i\ell_{i_1},1))$ ---or the arc $((\ell_i,\ell_i\overline{\ell}_i,6),(\ell_i,\ell_i\ell_{i_1},1))$ if $q=0$; next, the arcs $((\ell_i,\ell_i\ell_{i_1},6),(\ell_i,$ $\ell_i\ell_{i_2},1))$, $\ldots$, $((\ell_i,\ell_i\ell_{i_{s-1}},6),(\ell_i,\ell_i\ell_{i_s},1))$ and $((\ell_i,\ell_i\ell_{i_s},6),\alpha_{i+1})$, unless $i=n$, in which case we create $((\ell_i,\ell_i\ell_{i_s},6),\beta_1)$.

\begin{remark} \label{userem1} In the example of Figure~\ref{figCyHam3}, one can see that if a path takes, e.g., the arc $(\alpha_1,(x_1,x_1\overline{x}_1,1))$, then it visits the vertices $(x_1,x_1\overline{x}_1,6)$, $(x_1,x_1a_1,1)$, $(x_1,x_1a_1,6)$, and~$\alpha_2$. If on the other hand, we use the arc $(\alpha_1,(\overline{x}_1,x_1\overline{x}_1,1))$, we also go to~$\alpha_2$. The same is true between $\alpha_2$ and~$\alpha_3$, $\ldots$, between $\alpha_{n-1}$ and~$\alpha_n$, between $\alpha_n$ and~$\beta_1$.\end{remark}
We can see that so far, $\alpha_1$ has (out-)degree~2, $\alpha_2$, $\ldots$, $\alpha_n$ have degree~4 (in- and out-degrees equal to~2), and $\beta_1$ has (in-)degree~2.

\medskip

\noindent {\bf Step 3.} We consider the $m$ clauses and the $m$ corresponding triangles~$T_j$.

We create $2m$ vertices $\beta_j, \beta'_j$, $1\leq j \leq m$. As we have seen in the previous step, $\beta_1$ has already two in-neighbours, which can be $(\ell_n,\ell_nt_q,6)$, or $(\ell_n,\ell_n{\overline \ell}_n,6)$, or $(\ell_n,\ell_n\ell_{n_s},6)$. We also create one more vertex~$\delta$, which will have only in-neighbours, so that $\alpha_1$ and $\delta$ will necessarily be the ends of any directed Hamiltonian path, if such a path exists.

Now for the triangle $T_j=\{a_j,b_j,d_j\}$, $1\leq j \leq %m_c$,
m$, associated to the clause $c_j=\{\ell_{j_1},\ell_{j_2},\ell_{j_3}\}$ in the graph $G_{VC}$, we consider the six corresponding components $H_{a_jb_j}$, $H_{a_jd_j}$, $H_{b_jd_j}$, $H_{a_j\ell_{j_1}}$, $H_{b_j\ell_{j_2}}$ and $H_{d_j\ell_{j_3}}$. The vertices $\beta_j$ %(for $1\leq j \leq m_c$)
and $\beta'_j$ can be seen as triangle selectors, intended to choose two vertices among three. With this in mind, we create the following arcs (see Figure~\ref{figCyHam4}), for $j\in \{1,2,\ldots,%m_c\}$:
m\}$:

\begin{figure}
\begin{center}
\includegraphics*[scale=0.85]{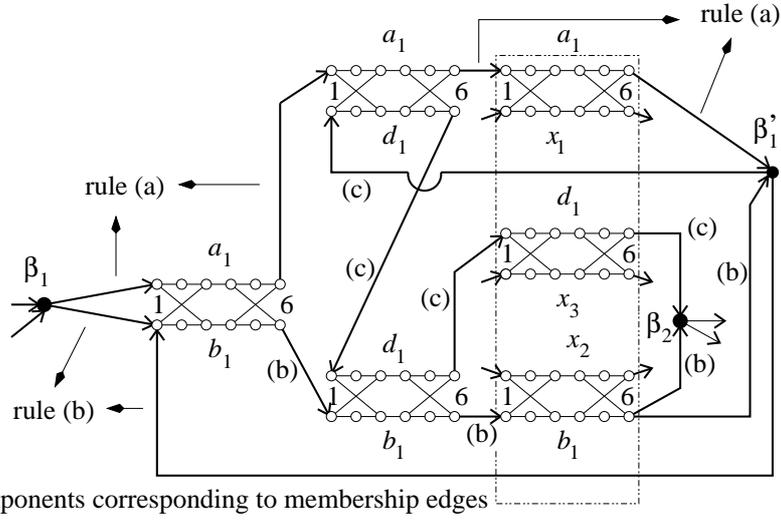}
\end{center}
\caption{The treatment of the triangle $T_1$ from Figure~\ref{figWC} (Step~3). The arrows inside $H_e$ are not represented.}
\label{figCyHam4}
\end{figure}
Rule (a): $(\beta_j,(a_j,a_jb_j,1))$, $((a_j,a_jb_j,6),(a_j,a_jd_j,1))$, $((a_j,a_jd_j,6),(a_j,a_j\ell_{j_1},$ $1))$, $((a_j,a_j\ell_{j_1},6),\beta'_j)$.

Rule (b): $(\beta_j,(b_j,a_jb_j,1))$, $(\beta'_j,(b_j,a_jb_j,1))$, $((b_j,a_jb_j,6),(b_j,b_jd_j,1))$, $((b_j,$ $b_jd_j,6),(b_j,b_j\ell_{j_2},1))$, $((b_j,b_j\ell_{j_2},6),\beta'_j)$, plus the arc $((b_j,b_j\ell_{j_2},6),\beta_{j+1})$, %including the case when $j=m_c$,
unless $j=m$, in which case it is $((b_j,b_j\ell_{j_2},6),\delta)$. 

Rule (c): $(\beta'_j,(d_j,a_jd_j,1))$, $((d_j,a_jd_j,6),(d_j,b_jd_j,1))$, $((d_j,b_jd_j,6),(d_j,d_j\ell_{j_3},$ $1))$, $((d_j,d_j\ell_{j_3},6),\beta_{j+1})$, unless $j=m$, in which case it is $((d_j,d_j\ell_{j_3},6),\delta)$. %IDEM (b)
\begin{remark} \label{remselect} In the example of Figure~\ref{figCyHam4}, there are three ways for going from $\beta_1$ to $\beta_2$ through the components $H_{a_1b_1}$, $H_{a_1d_1}$ and~$H_{d_1b_1}$. %%%, 

\medskip  

\noindent $\bullet$ If a path starts by taking the arc $(\beta_1,(a_1,a_1b_1,1))$, then there are two possibilities, according to how we visit~$H_{a_1b_1}$:

$\circ$ The first possibility corresponds to taking $a_1$ and $d_1$, not $b_1$, in a vertex cover: the path completely visits the component $H_{a_1b_1}$ from the $a_1$-side, then the component $H_{a_1d_1}$ in parallel, then the component $H_{a_1x_1}$ in a so far unspecified way, then~$\beta_1'$.

Next, it takes the arc $(\beta'_1,(d_1,d_1a_1,1))$, re-visits $H_{d_1a_1}$ in parallel, completely visits $H_{d_1b_1}$ from the $d_1$-side, then $H_{d_1x_3}$, and ends this path section at~$\beta_2$. One can see that the three components corresponding to edges incident to~$b_1$ must all be completely visited from the side opposite~$b_1$, including the $x_2$-side.

$\circ$ The second possibility corresponds to taking $a_1$ and $b_1$, not $d_1$, in a vertex cover: the path follows the arc $(\beta_1,(a_1,a_1b_1,1))$, visits $H_{a_1b_1}$ in parallel, visits completely $H_{a_1d_1}$ from the $a_1$-side, then $H_{a_1x_1}$, and~$\beta_1'$. %%% (if the path goes instead to~$\beta_2$, then either it gets stuck or it leaves without visiting all the vertices).

Next, it takes the arc $(\beta'_1,(b_1,b_1a_1,1))$, goes through $H_{b_1a_1}$ in parallel, goes completely through $H_{b_1d_1}$ from the $b_1$-side, then $H_{b_1x_2}$, and ends this path section at~$\beta_2$. The component $H_{d_1x_3}$ is not yet visited.

\medskip

\noindent $\bullet$ Alternatively, a path can start by taking the arc $(\beta_1,(b_1,a_1b_1,1))$; this corresponds to taking $b_1$ and~$d_1$, not~$a_1$, in a vertex cover and constitutes the third way for going from $\beta _1$ to $\beta _2$. The path then completely visits $H_{b_1a_1}$ from the $b_1$-side, $H_{b_1d_1}$ in parallel, $H_{b_1x_2}$, and~$\beta'_1$.

Next, it completely visits $H_{d_1a_1}$ from the $d_1$-side, $H_{d_1b_1}$ in parallel, $H_{d_1x_3}$, and this path section ends at~$\beta_2$. The component $H_{a_1x_1}$ is not yet visited.

\medskip

\noindent It is easy to see that these are the {\rm only} three ways for going from $\beta_1$ to $\beta_2$ through the components $H_{a_1b_1}$, $H_{a_1d_1}$ and~$H_{d_1b_1}$, not taking into account the ways of going through the components $H_{a_1x_1}$, $H_{b_1x_2}$ and $H_{d_1x_3}$ (this issue will be treated later on, in the general case): indeed, the only possibility left would be to follow the arc $(\beta_1,(b_1,a_1b_1,1))$ and visit $H_{b_1a_1}$ in parallel, but then the $a_1$-side of $H_{b_1a_1}$ cannot be reached.

\medskip  

\noindent The same will be true for the components $H_{a_jb_j}$, $H_{a_jd_j}$ and~$H_{d_jb_j}$ and the corresponding triangles $T_j$, $1\leq j \leq %m_c$,
m$, between $\beta_j$ and $\beta_{j+1}$ %(or $\beta_{m_c}$ and~$\gamma_1$ when $m_c=m$).
(or between $\beta_{m}$ and~$\delta$).%NON plutot $\beta_{m_c+1}$ sauf si $m_c=m$ {\bf A VOIR}
\end{remark}
The description of the oriented graph~$H$ is complete. Now $\beta_1$ has increased its degree to~4, and $\beta_2$, $\ldots$, %$\beta_{m_c}$
$\beta_m$ and $\beta'_1$, $\ldots$, %$\beta'_{m_c}$
$\beta'_m$ have degree~4. {\it Actually, all the selectors but }$\alpha_1$ {\it have in-degree}~2 {\it and out-degree}~2 {\it in}~$H$. These $n$ selectors $\alpha_i$, $1\leq i \leq n$, %%% $2m_c$
and $2m$ selectors $\beta_j$, $\beta_j'$, $1\leq j \leq %%% m_c$,
m$, %%% and $m_F$ selectors $\beta_j$, $m_c+1 \leq j \leq m$,
translate the choices we have to make when constructing a vertex cover with size $2m+n$: we have one choice among the $n$ variables (take $x_i$ or~${\overline x}_i$); %%% , one choice in the $m_F$ triangles $T_j=\{a_j,b_j,d_j\}$ with $d_j$ linked to~$F$ (take $a_j$ or~$b_j$);
as for the $m$~triangles~$T_j$ associated to the clauses, %%% without~F,
Remark~\ref{remselect} has shown how the selectors $\beta_j$, $\beta_j'$, $1\leq j \leq %%% m_c$,
m$, can be used to choose two vertices among three. The {\it number} of selectors is one reason why there is no directed Hamiltonian path in~$H$ when the vertex covers in $G_{VC}$ have size at least $2m+n+1$.

The order of $H$ is $12|E_{VC}|+n+2m+1$, which is at most $12(n+9m)+n+2m+1$, so that the transformation is polynomial indeed.

\medskip

\noindent We are now going to prove our claims about the existence or non-existence, uniqueness or non-uniqueness, of a directed Hamiltonian path in~$H$.

\medskip

\noindent {\bf C) How it Works}\\
Assume first that there is an assignment satisfying the instance of U-1-3-SAT, and therefore that there is a vertex cover $V^*$ in $G_{VC}$ with size $2m+n$. We construct a path in~$H$ in a straightforward way: every component $H_{uv}$ ($uv\in E_{VC}$) with $\{u,v\} \subset V^*$ is visited in parallel, whereas $H_{uv}$ is completely visited from the $u$-side whenever $u\in V^*$, $v\notin V^*$. Let us have a closer look at how this works:

We start at $\alpha_1$, and visit completely the component $H_{x_1{\overline x}_1}$ from the $x_1$-side if $x_1=\,$T, from the ${\overline x}_1$-side if $x_1=\,$F (or, equivalently, if $x_1\in V^*$ or ${\overline x}_1\in V^*$, respectively). If, say, $x_1=\,$F, we then completely go through all the components corresponding to triangles~$T_j$ and involving ${\overline x}_1$, all from the ${\overline x}_1$-side;  note that all the components just completely visited involve ${\overline x}_1$ and a vertex not in~$V^*$, by the very construction of the vertex cover~$V^*$, which is possible because it stems from an assignment 1-3-satisfying all the clauses. Then we go through the components constructed from the edge sets $E'_j$ %%% and $E'_k$
and involving~${\overline x}_1$; those involving a second vertex in~$V^*$ (i.e., a true literal) are visited in parallel, whereas those involving a vertex not in~$V^*$ are completely visited from the ${\overline x}_1$-side; then the path arrives at~$\alpha_2$. The components involving~$x_1$, apart from~$H_{x_1{\overline x}_1}$, remain completely unvisited for the time being, and the components that have been visited in parallel will have to be re-visited.

We act similarly between $\alpha_2$ and $\alpha_3$, $\ldots$, $\alpha_n$ and~$\beta_1$; cf. Remark~\ref{userem1}. When doing this, we re-visit all the components that had been visited only in parallel, and completely visit the components involving a literal not in~$V^*$ and corresponding to edges in~$E'_j$. %%% and~$E'_k$.
The only components not visited yet between $\alpha_1$ and~$\beta_1$ are those corresponding to edges between a false literal (not in~$V^*$) and its neighbours in the triangles~$T_j$.

Next, starting from $\beta_1$, we use Remark~\ref{remselect} according to the three possible cases: (a)~$\{a_1,d_1\}\subset V^*,$ $b_1\notin V^*$, (b)~$\{a_1,b_1\}\subset V^*,$ $d_1\notin V^*$, (c)~$\{b_1,d_1\}\subset V^*,$ $a_1\notin V^*$. We give in detail only the third case, for the clause $c_1=\{\ell_1,\ell_2,\ell_3\}$: we use the arc $(\beta_1,(b_1,a_1b_1,1))$ and completely visit the component $H_{a_1b_1}$ from the $b_1$-side, then the component $H_{b_1d_1}$ in parallel, then the complete component $H_{b_1\ell_2}$ from the $b_1$-side (because if $b_1\in V^*$, then $\ell_2\notin V^*$ and this component had not yet been visited) and end at~$\beta'_1$. Next, we take the arc $(\beta'_1,(d_1,d_1a_1,1))$, we completely visit $H_{d_1a_1}$ from the $d_1$-side, re-visit $H_{d_1b_1}$ in parallel, completely visit $H_{d_1\ell_3}$ from the $d_1$-side, and this path section ends at~$\beta_2$. Note that (a)~the three components involving~$a_1$ between $\beta_1$ and~$\beta_2$ have been completely visited, from the $b_1$-, $d_1$- or~$\ell_1$-sides (because $a_1 \notin V^*$ implies that $\ell_1 \in V^*$); (b)~any so far unvisited component involving a false literal (here, these are $\ell_2$ and~$\ell_3$) and one of the vertices of the triangle~$T_1$ (here $b_1$ and~$d_1$) has now been completely visited from the triangle sides (here from the $b_1$- and $d_1$-sides).

We act similarly between $\beta_2$ and $\beta_3$, $\ldots$, %%% $\beta_{m_c}$
$\beta_m$ and~%$\beta_{m_c+1}$. If $m_c=m$, i.e., $m_F=0$, instead of $\beta_{m_c+1}$, we end this cycle section at~
$\delta$; cf. the end of Remark~\ref{remselect}. %%%
The ultimate section takes us between $\beta_m$ and~$\delta$, the final vertex, and we have indeed built a directed Hamiltonian path, from~$\alpha_1$ to~$\delta$, in the oriented graph~$H$.

\medskip

\noindent Obviously, two different assignments 1-3-satisfying all the clauses lead, following the above process, to two different paths in~$H$. We still want to prove that 1)~if no assignment 1-3-satisfying all the clauses exists, then no path exists, and 2)~a unique assignment 1-3-satisfying all the clauses leads to a unique path.

1) We assume that there is a directed Hamiltonian path ${\cal HP}$ in~$H$, and exhibit an assignment 1-3-satisfying all the clauses.

Let us consider the vertex~$\alpha_1$; its two out-neighbours in~$H$ are $(x_1,x_1\overline{x}_1,1)$ and $(\overline{x}_1,x_1\overline{x}_1,1)$. 
%Assume first that the two neighbours of~$\alpha_1$ in~${\cal HC}$ are $(b_m,b_m\ell_{j_2},6)$ and $(d_m,d_m\ell_{j_3},6)$: at $(d_m,d_m\ell_{j_3},6)$, we have no choice and must go, through $H_{d_m\ell_{j_3}}$, $H_{d_mb_m}$ and $H_{d_ma_m}$, to~$\beta'_m$; from $(b_m,b_m\ell_{j_2},6)$, we can only go to~$\beta_m$; then from~$\beta'_m$ we can go nowhere without being stuck in a smaller cycle, containing $\beta_m$ and~$\beta'_m$. If, on the other hand, the two neighbours of~$\alpha_1$ in~${\cal HC}$ are $(x_1,x_1\overline{x}_1,1)$ and $(\overline{x}_1,x_1\overline{x}_1,1)$, then necessarily we visit $H_{x_1\overline{x}_1}$ in parallel, and then, see Remark~\ref{userem1} and Figure~\ref{figCyHam3}, both parts of the cycle end at~$\alpha_2$ and we cannot go further: we are stuck in a small cycle containing $\alpha_1$ and~$\alpha_2$.
So exactly one of the arcs $(\alpha_1,(x_1,x_1\overline{x}_1,1))$, $(\alpha_1,(\overline{x}_1,x_1\overline{x}_1,1))$ is part of~${\cal HP}$. %The same argument can be used {\it mutatis mutandis} for $\alpha_i$, $1<i<n$, (using $\alpha_i$ and $\alpha_{i+1}$ or $\alpha_{i-1}$ and~$\alpha_i)$ and for~$\alpha_n$ (using $\alpha_n$ and $\beta_1$ or $\alpha_{n-1}$ and~$\alpha_n$).
The same is true for $\alpha_i$, $1<i\leq n$. As a consequence, we can define a valid assignment of the variables~$x_i$, $1\leq i \leq n$, by setting $x_i=\,$T if and only if the arc $(\alpha_i,(x_i,x_i\overline{x}_i,1))$ belongs to~${\cal HP}$.

Next, we address the vertices~$\beta_j$, $1\leq j\leq m$. The construction in Steps~2 and~3 is such that %:\\
%$\bullet$ The vertex $\beta_1$ is linked to four vertices: $(a_1,a_1b_1,1)$, $(b_1,a_1b_1,1)$, and two among $(\ell_n,\ell_nt_q,6)$, $(\ell_n,\ell_n{\overline \ell}_n,6)$ and $(\ell_n, \ell_n\ell_{n_s},6)$.\\
%$\bullet$ Each
each vertex $\beta_j$, $1\leq j\leq m$, has two out-neighbours, $(a_j,a_jb_j,1)$ and $(b_j,a_jb_j,1)$. %, $(b_{j-1},b_{j-1}\ell_{(j-1)_2},6)$ and $(d_{j-1},d_{j-1}\ell_{(j-1)_3},6)$.
 
%For all the vertices $\beta_j$, $1\leq j \leq m$, using the same type of argument as previously, we can show that in~${\cal HC}$, the vertex $\beta_j$ has one neighbour with a~1 in third place, and one neighbour with a~6 in third place.
This implies that the assignment defined above is such that there is at least one true literal in each clause. Indeed, if we assume that the clause $c_j=\{\ell_{j_1},\ell_{j_2},\ell_{j_3}\}$ does not contain any true literal, then the component $H_{a_j\ell_{j_1}}$ is completely visited by~${\cal HP}$ from the $a_j$-side, because $\ell_{j_1}=\,$F implies that the arc $(\alpha_j,(\ell_{j_1},\ell_{j_1}{\overline \ell}_{j_1},1))$ is not part of~${\cal HP}$ and does not give access to the $\ell_{j_1}$-side. Similarly, the components $H_{b_j\ell_{j_2}}$ and $H_{d_j\ell_{j_3}}$ are completely visited by~${\cal HP}$ from the $b_j$- and $d_j$-sides, respectively. This in turn implies that in~${\cal HP}$ we have the arcs $((a_j,a_j\ell_{j_1},6),\beta'_j)$, $(\beta_j,(a_j,a_jb_j,1))$, $((d_j,d_j\ell_{j_3},6),\beta_{j+1})$ and $(\beta'_j,(d_j,d_ja_j,1))$ ---replace $\beta_{j+1}$ by $\delta$ if $j=m$. Now how does ${\cal HP}$ go through~$(b_j,b_j\ell_{j_2},6)$? It cannot be with the help of the $\ell_{j_2}$-side of $H_{b_j\ell_{j_2}}$, so there are only two possibilities left: but if it is with the arc $((b_j,b_j\ell_{j_2},6),\beta'_j)$, then $\beta'_j$ has three neighbours in~${\cal HP}$, which is impossible; and if it is with the arc $((b_j,b_j\ell_{j_2},6),\beta_{j+1})$, then in~${\cal HP}$, the vertex $\beta_{j+1}$ has two in-neighbours, which is impossible ---including when $j=m$ and $\beta_{j+1}$ is replaced by~$\delta$. {F}rom this we can conclude that the clause $c_j=\{\ell_{j_1},\ell_{j_2},\ell_{j_3}\}$ contains at least one true literal. 

Assume next that one clause has at least two true literals: without loss of generality, $c_j=\{\ell_{j_1},\ell_{j_2},\ell_{j_3}\}$ is such that $\ell_{j_1}=\ell_{j_2}=\,$T. %%% (this can include the case when F appears in the clause).
Then ${\cal HP}$ has no access to the ${\overline \ell}_{j_1}$- and ${\overline \ell}_{j_2}$- sides of the components involving ${\overline \ell}_{j_1}$ or ${\overline \ell}_{j_2}$, but, since there is the edge ${\overline \ell}_{j_1}{\overline \ell}_{j_2}$ in $G_{VC}$, this means that ${\cal HP}$ has no way of visiting the component $H_{{\overline \ell}_{j_1}{\overline \ell}_{j_2}}$. Therefore, we have just established that the assignment derived from the path~${\cal HP}$ 1-3-satisfies all the clauses. %%%By contraposition, if there is no assignment 1-3-satisfying the clauses, there is no Hamiltonian cycle.
This, together with the fact that two assignments 1-3-satisfying the clauses lead to two paths, shows that a NO answer to the instance of U-1-3-SAT implies a NO answer for the constructed instance~$H$ of U-HAMP[O].

2) We want to show that a unique assignment~${\cal A}$ 1-3-satisfying all the clauses leads to a unique path in~$H$. This assignment leads to a unique vertex cover $V^*$, of size $n+2m$, in~$G_{VC}$, and to a path in~$H$, as already seen. Now assume that we have a second path, so that these two paths, which we call ${\cal HP}_1$ and~${\cal HP}_2$, both lead, with the above description in~1), to the same~${\cal A}$ and the same~$V^*$.

The two paths must behave in the same way over the components $H_{x_i{\overline x}_i}$, $1\leq i \leq n$: otherwise, from them we could define two different valid assignments, which would both, as seen previously, 1-3-satisfy the clauses.
%%%
%%% MODIF

Next, consider the clause $c_j=\{\ell_{j_1}, \ell_{j_2}, \ell_{j_3}\}$ and assume without loss of generality that ${\cal A}(\ell_{j_1})=\,$T, ${\cal A}(\ell_{j_2})={\cal A}(\ell_{j_3})=\,$F; this implies, for both ${\cal HP}_1$ and~${\cal HP}_2$, that the components $H_{\ell_{j_1}{\overline \ell}_{j_1}}$, $H_{\ell_{j_2}{\overline \ell}_{j_2}}$ and $H_{\ell_{j_3}{\overline \ell}_{j_3}}$ are completely visited from the $\ell_{j_1}$-, ${\overline \ell}_{j_2}$- and ${\overline \ell}_{j_3}$-sides, respectively, so that both paths have no access to the $\ell_{j_2}$- nor $\ell_{j_3}$-sides. As a consequence, between $\beta_j$ and $\beta_{j+1}$ (or $\beta_m$ and~$\delta$), the components $H_{b_j\ell_{j_2}}$ and $H_{d_j\ell_{j_3}}$ are completely visited from the $b_j$- and $d_j$-sides, respectively. Then necessarily the following arcs belong to ${\cal HP}_1$ and~${\cal HP}_2$, going along the $d_j$-side:

\noindent $((d_j,d_j\ell_{j_3},6),\beta_{j+1})$ ---or $((d_j,d_j\ell_{j_3},6),\delta)$---, $((d_j,d_jb_j,6),(d_j,d_j\ell_{j_3},1))$, $((d_j,d_ja_j,$ $6),(d_j,d_jb_j,1))$, $(\beta'_j,(d_j,d_ja_j,1))$;

and going along the $b_j$-side:

\noindent $((b_j,b_j\ell_{j_2},6),\beta'_j)$ (because $\beta_{j+1}$ ---or~$\delta$--- cannot have two in-neighbours), and $((b_j,b_jd_j,6),(b_j,b_j\ell_{j_2},1))$, $((b_j,b_ja_j,6),(b_j,b_jd_j,1))$.

The component $H_{b_jd_j}$ must be visited in parallel, and it is $(\beta_j,(b_j,b_ja_j,1))$ that belongs to the two paths.

We can see that all the components containing~$a_j$, in particular $H_{a_j\ell_{j_1}}$, must be completely visited from the sides opposite~$a_j$. So far, we have proved that the two paths ${\cal HP}_1$ and~${\cal HP}_2$ behave identically between $\beta_j$ and $\beta_{j+1}$ (or $\beta_m$ and~$\delta$), including on the components corresponding to membership edges (between literals and triangles).

Consider now what happens between $\alpha_i$ and $\alpha_{i+1}$ (or $\alpha_n$ and~$\beta_1$). Assume without loss of generality that, say, ${\cal A}(x_i)=\,$T, so that $(\alpha_i,(x_i,x_i{\overline x}_i,1))$ is part of the two paths. Consider the components involving~$x_i$ in~$H$: there are first those involving vertices of type $a$, $b$ or~$d$, which translate the membership of $x_i$ to a certain number of clauses, and which we called $t_1,\ldots ,t_q$ in Step~2(a); we have already seen in the previous paragraph that these  components must be completely visited from the $x_i$-side.

Then we consider the components created from the edges in~$E'_j$, cf. Step~2(b); here, some edges in $G_{VC}$ can have both ends in~$V^*$, but, using similar arguments as before, we can see that the two paths will visit all these components in the same way: consider the clause $c_j=\{\ell_{j_1}, \ell_{j_2}, \ell_{j_3}\}$ and the corresponding set~$E'_j$, and assume without loss of generality that $x_i=\ell_{j_1}$, so that ${\cal A}(\ell_{j_1})=\,$T, which implies that ${\cal A}(\ell_{j_2})={\cal A}(\ell_{j_3})=\,$F; then $H_{{\overline \ell}_{j_1}{\overline \ell}_{j_2}}$ and $H_{{\overline \ell}_{j_1}{\overline \ell}_{j_3}}$ must be completely visited from the ${\overline \ell}_{j_2}$- and ${\overline \ell}_{j_3}$-sides, respectively, and $H_{{\overline \ell}_{j_2}{\overline \ell}_{j_3}}$ in parallel, i.e., the two paths have no choice but to behave identically on all three components. As for the components with ${\overline x}_i$, they must be completely visited from the side which is not the side of~${\overline x}_i$.

So we have just proved that the two paths are identical between $\alpha_1$ and $\alpha_2$, $\ldots$, $\alpha_n$ and~$\beta_1$. 
%The proof is pretty much the same for the components lying between $\beta_1$ and $\beta_2$, $\ldots$, up to~$\alpha_1$: consider the clause $c_j=\{\ell_{j_1}, \ell_{j_2}, \ell_{j_3}\}$ and assume without loss of generality that ${\cal A}(\ell_{j_1})=\,$T, i.e., $a_j \notin V^*$, $b_j\in V^*$, $d_j\in V^*$. Between $\beta_j$ and~$\beta_{j+1}$ (or~$\alpha_1$), we have seen previously that the component $H_{a_j \ell_{j_1}}$ must be completely visited from the $\ell_{j_1}$- side only, and the components $H_{b_j \ell_{j_2}}$ and $H_{d_j \ell_{j_13}}$ from the $b_j$- and $d_j$-sides only, respectively. Then there is no choice left for visiting the vertices between $\beta_j$ and~$\beta_{j+1}$ (or~$\alpha_1$). Indeed, any path must go from $(d_j,d_j\ell_{j_3},6)$ to~$\beta_{j+1}$ (or~$\alpha_1$), from $(d_j,d_j\ell_{j_3},1)$ to $(d_j,d_jb_j,6)$, from $(b_j,b_j\ell_{j_2},1)$ to $(b_j,b_jd_j,6)$; then $H_{d_jb_j}$ must be visited in parallel; then any path must go from $(b_j,b_j\ell_{j_2},6)$ to~$\beta'_j$, because $\beta_{j+1}$ (or~$\alpha_1$) cannot have two neighbours with third position equal to~6; the rest is easily seen to be forced.

Therefore, the two paths (between $\alpha_1$ and~$\delta$) are one and the same. \qed}
% {\bf The Travelling Salesman ?????}

\pagebreak

\end{document}